\documentclass[twocolumn,showpacs,amsmath,amssymb,aps,prd]{revtex4}
\usepackage{amsmath,amsfonts,bm}
\usepackage{graphicx}
\usepackage[colorlinks, linkcolor={red},citecolor={blue}]{hyperref}
\newcommand{\mc}[1]{\mathcal{#1}}
\newcommand{\f}[2]{\frac{#1}{#2}}
\begin{document}
\title{Anisotropy in Born-Infeld brane cosmology}
\author{Zahra Haghani}
\email{z$\_$haghani@sbu.ac.ir}
\author{Hamid Reza Sepangi}
\email{hr-sepangi@sbu.ac.ir}
\author{Shahab Shahidi}
\email{s$\_$shahidi@sbu.ac.ir} \affiliation{Department of Physics,
Shahid Beheshti University, G. C., Evin,Tehran 19839, Iran}
\begin{abstract}
The accelerated expansion of the universe together with its present
day isotropy has posed an interesting  challenge to the numerous
model theories presented over the years to describe them. In this
paper, we address the above questions in the context of a
brane-world model where the universe is filled with a Born-Infeld
matter. We show that in such a model, the universe evolves from a
highly anisotropic state to its present isotropic form which has
entered an accelerated expanding phase.
\end{abstract}
\maketitle

\section{Introduction}
The problem of the diverging Coulomb field and self-energy of point
particles in Maxwell's theory of electrodynamics led Born and Infeld
(BI) in  1934 to the construction of a nonlinear extension of
classical electrodynamics \cite{1}. In recent years, the study and
use of nonlinear electrodynamics has been the focus of much
attention in the context of cosmology. Such strong motivations stem
from developments in string/M-theory, where it has been shown that
the BI theory naturally arises in the low energy limit of the open
string theory \cite{2}. Nonlinear electrodynamics has also been
applied to several branches of physics, namely, the effective
theories at different levels of string/M-theory \cite{3},
cosmological models \cite{4,4.5}, black holes \cite{5,6} and wormhole
physics \cite{7}, among others.

The ubiquitous Friedmann-Robertson-Walker (FRW) model has long been
considered as a standard against which other model theories are
compared. In this model the universe is regarded as homogeneous and
isotropic with only one dynamical feature, its rate of expansion or
contraction. Its high degree of symmetry makes it easy to analyse
but also makes it somewhat unrealistic. However, it does model the
average properties of the observed universe, on the largest scales,
quite well.  The Bianchi models, on the other hand, pose a more
realistic scenario. These models represent a homogeneous but
anisotropic universe which  are essentially indistinguishable from
FRW models far from the initial singularity or, in the case of a
collapsing universe, far from the future singularity. The present
epoch is far from all such singularities. Bianchi models have three
dynamical parameters, the expansion or contraction in the three
spatial directions which are, in general, all different. The three
rates of expansion or contraction asymptotically (in time) equalize,
thus producing a model which looks like that of a FRW. If the
universe did not begin by expanding exactly isotropically, then the
Bianchi models would describe its early behavior more accurately
than the FRW model.

The idea that our four-dimensional ($4D$) universe is a hypersurface
(brane) in a $5D$ spacetime (bulk) \cite{8, 9, 10} has been the
dominant idea behind many model theories attempting to explain the
observed universe over the past decade. One of the most successful
of such higher dimensional models is that proposed by Randall and
Sundrum where the bulk has the geometry of an AdS space admitting
$\mathbb{Z}_2$ symmetry  \cite{9}. They were successful in
explaining what is known as the hierarchy problem; the enormous
disparity between the strength of the fundamental forces. The
Randall-Sundrum (RS) scenario has had a great impact on our
understanding of the universe and has brought higher dimensional
gravitational theories to the fore. In certain RS type models, all
matter and gauge interactions reside on the brane while gravity can
propagate into the bulk. In obtaining the field equations on the brane one uses the Israel junction
conditions \cite{11} and the Gauss-Codazzi equations, as employed by
Shiromizu, Maeda and Sasaki (SMS) \cite{12}. These field equations
differ from the standard Einstein field equations in $4D$ in that
they have additional terms like $\pi_{\mu\nu}$ which depend on the
energy-momentum tensor on the brane and the electric part of the
Weyl tensor  $\mc{E}_{\mu\nu}$, leading to the appearance of a
quadratic term in the Friedmann equation. This term was initially
considered as a possible solution to the accelerated expansion of
the universe. However, soon it was realized to be incompatible with
the big bang nucleosynthesis, requiring additional fixes \cite{13}.

Since their introduction, the RS type models have been employed
frequently in conjunction with a myriad of matter fields to study
the behavior of the observed universe. What has been lacking however
is such a study with non-linear electrodynamics of the BI as the
matter field. However, as we will show in the Appendix, the BI
matter cannot be used in a brane-world scenario together with Israel
junction conditions. The reason is that the solutions of the
resulting field equations predict a universe which is incompatible
with present day observations. In addition, there have been
arguments on the uniqueness of the junction conditions \cite{16} or
their use when more than one non-compact extra dimension is
involved. In view of the above, one is  lead to consider an
alternative brane-world scenario in which the Israel junction
conditions become redundant. Brane-world scenarios under more
general conditions and still compatible with the brane-world program
have therefore been rather extensively studied over the recent past
where it has been shown that it is possible to find a set of
cosmological solutions in accordance with the current observations.
Under these conditions, without using $\mathbb{Z}_2$ symmetry or
postulating any junction conditions, the Friedmann equation is
modified by a geometrical term which is defined in terms of the
extrinsic curvature, leading to a geometrical interpretation of dark
energy \cite{15}. An example of such a scenario was presented in
\cite{17} where particles are trapped on a $4D$ hypersurface by the
action of a confining potential. The dynamics of test particles
confined to a brane by the action of such a potential at the
classical and quantum levels were studied in \cite{18}. In \cite{14}
the authors showed that the quadratic term in the energy-density can
be produced in this model, suggesting that the model is identical to
the SMS method at least in the isotropic universe. In this paper, we
consider an anisotropic brane-world  with Bianchi type I geometry
filled with a nonlinear electromagnetic field of the BI type. The
electromagnetic tensor has six independent components of which only
two survive to construct the BI energy-momentum tensor. The
anisotropy parameter and the deceleration parameter are calculated
and shown to be compatible with present observations, namely that
the universe is in an accelerating phase and expanding
isotropically.

\section{The setup}
Let us start by defining the Bianchi type I universe through the
metric
\begin{align}
 ds^2=-dt\otimes dt+\sum_{i=1}^3 a_i(t)^2 dx^i\otimes
dx^i, \label{line}
\end{align}
where $a_i$ are the scale factors. We also have the usual definition
of the volume scale factor $v$, the directional Hubble parameters
$H_i$ and the mean Hubble parameter $H$, defined as
\begin{align}
v=\prod_{i=1}^3 &a_i, \qquad H_i=\frac{\dot{a}_i}{a_i},\qquad
3H=\sum_{i=1}^3 H_i,\nonumber\\
&\Delta H_i=H_i-H.\qquad i=1,2,3.
\label{def1}
\end{align}
 The physical observables are the anisotropy parameter
$A$ and the deceleration parameter  $q$ given by
\begin{align}
 3A=\sum_{i=1}^3\left(\frac{\Delta H_i}{H}\right)^2,\qquad
q=\frac{d}{dt} H^{-1}-1. \label{def2}
\end{align}

In view of the discussion above, we focus attention on a brane world
scenario \cite{18} in which the use of $\mathbb{Z}_2$ symmetry and
Israel junction conditions are relaxed in favor of a confining
potential $\mathcal{V}$ whose role is to localize the gauge fields
of the standard model on the brane. In this scenario a new conserved
quantity, $Q_{\mu\nu}$, appears which is totally geometric in nature
and depends on the extrinsic curvature only, reflecting the effects
of the extra dimension. In the  SMS method one replaces this tensor
with a tensor build up by the energy-momentum tensor using the
Israel junction conditions. We will compute this tensor using the
Codazzi equation. The field equation of the model is given by
\cite{19}
\begin{align}
 G_{\mu\nu}=\alpha\tau_{\mu\nu}-\Lambda
g_{\mu\nu}+Q_{\mu\nu}+\mc{E}_{\mu\nu},\label{eq1.1}
\end{align}
where $\alpha$ is the energy scale of the brane, $\Lambda$ is the
cosmological constant, $\mc{E}_{\mu\nu}$ is the electric part of the
Weyl tensor and $Q_{\mu\nu}$ is a conserved quantity, $\nabla_\mu
Q^{\mu\nu}=0$,  expressed in terms of the extrinsic curvature
$K_{\mu\nu}$ and its trace
\begin{align}
 Q_{\mu\nu}=KK_{\mu\nu}-K_{\mu\alpha}K^{\alpha}_{\nu}+\f{1}{2}
\left(K_{\alpha\beta}K^{\alpha\beta}-K^2\right)g_{\mu\nu}\label{ext}.
\end{align}
In order to calculate $Q_{\mu\nu}$ we first use the York's relation
\begin{align}
 K_{\mu\nu}=-\f{1}{2}\f{\partial g_{\mu\nu}}{\partial \xi},
\end{align}
where $\xi$ is a vector normal to the brane. This tells us that the
extrinsic curvature is diagonal and does not depend on the spatial
coordinates. After separating the spatial components, the Codazzi
equation relates the diagonal elements of the extrinsic curvature
tensor \cite{19}
\begin{align}
&K^i_{~j,k}=K^i_{~k,j},\qquad i,j,k=1,2,3,\nonumber\\
&K^i_{~i,0}+\f{\dot{a}_i}{a_i}K^i_{~i}=\f{\dot{a}_i}{a_i}K^0_{~0}.\qquad(\textmd{no summation})\label{KK}
\end{align}
The first equation simply emphasizes the result that the spatial
components of the extrinsic curvature do not depend on the spatial
coordinates. The choice of the ansatz $K_{00}=d(t)$ for the temporal
component of the extrinsic curvature  and use of the second equation
in (\ref{KK}) leads one to the other components of the extrinsic
curvature
\begin{align}
K_{ii}=\left(c_i-e_i a_i\right)a_i,
\end{align}
where
\begin{align}
 e_i(t)=\f{1}{a_i(t)}\int d(t)\dot{a}_i(t)\textmd{d}t,
\end{align}
with $c_i$'s being constants. Using equation (\ref{ext}) we can
obtain the components of the tensor $Q^{\mu}_{~\nu}$
\begin{align}
Q^0_{~0}=&-(e_1e_2+e_2e_3+e_3e_1)\nonumber\\
&+\left(\f{c_1}{a_1}(e_2+e_3)+\f{c_2}{a_2}(e_1+e_3)+\f{c_3}{a_3}(e_2+e_3)\right)
\nonumber\\&-\left(\f{c_1}{a_1}\f{c_2}{a_2}+\f{c_2}{a_2}\f{c_3}{a_3}+
\f{c_3}{a_3}\f{c_1}{a_1}\right),\label{ext00}
\end{align}
\begin{align}
 Q^1_{~1}=&-e_2e_3-d(e_2+e_3)+\left(e_2\f{c_3}{a_3}+e_3\f{c_2}{a_2}\right)\nonumber\\
&+d\left(\f{c_2}{a_2}+\f{c_3}{a_3}\right)-\f{c_2}{a_2}\f{c_3}{a_3}.
\label{ext11}
\end{align}
The components $Q^2_{~2}$ and $Q^3_{~3}$ are obtained from
(\ref{ext11}) by permutation of indices. Let us now concentrate on
the case  where the temporal component of the extrinsic curvature is
constant, taking it as $d(t)=1$. Such a choice can be understood  in
terms of the definition of $K_{\mu\nu}$ and  observation that
$g_{00}=-1$. In this case the components of $Q^{\mu}_{~\nu}$ take
the simpler form
\begin{align}
 Q^0_{~0}=&-3+2\left(\f{c_1}{a_1}+\f{c_2}{a_2}+\f{c_3}{a_3}\right)\nonumber\\
&-\left(\f{c_1}{a_1}\f{c_2}{a_2}+\f{c_2}{a_2}\f{c_3}{a_3}+\f{c_3}{a_3}\f{c_1}{a_1}\right),\nonumber\\
Q^1_{~1}=&-3+2\left(\f{c_2}{a_2}+\f{c_3}{a_3}\right)-\f{c_2}{a_2}\f{c_3}{a_3},\label{ext22}
\end{align}
and again $Q^2_{~2}$ and $Q^3_{~3}$ are obtained by permutating the
indices.  If we assume that our $5D$ bulk has a constant curvature,
$\mc{E}_{\mu\nu}$ vanishes and the field equation becomes
\begin{align}
G_{\mu\nu}=\alpha\tau_{\mu\nu}-\Lambda g_{\mu\nu}+Q_{\mu\nu}.\label{eq1}
\end{align}

Let us now assume that the $4D$ universe is filled with the BI field
and is described by the following action
\begin{align}
 S_F=\int d^4 x \mathcal{L}(F), \label{lag1}
\end{align}
where
\begin{align}
 \mathcal{L}(F)&=\sqrt{-g}L(F)\nonumber\\
&=\f{\beta^2}{4\pi}\left\{
\sqrt{-g}-\sqrt{\mid\textmd{det}\left(g_{\mu\nu}+\beta^{-1}F_{\mu\nu}\right)\mid}\right\}.
\label{lag3}
\end{align}
Here, $\beta$ is a coupling constant with the dimension of the
field. In four space-time dimensions equation (\ref{lag3}) can be
expanded to give \cite{20}
\begin{align}
 L(F)=\f{\beta^2}{4\pi}\left(1-\mc{R}\right),
\label{lag2}
\end{align}
where
\begin{align}
\mc{R}=\sqrt{1+\frac{1}{2\beta^2}F^{\rho\sigma}F_{\rho\sigma}-
\frac{1}{16\beta^4}\left(\tilde{F}^{\rho\sigma}
F_{\rho\sigma}\right)^2}.\label{R}
\end{align}
In this equation, $F_{\mu\nu}=A_{\mu,\nu}-A_{\nu,\mu}$ is the usual
electromagnetic tensor with 4-potential $A^\mu$ and
$\tilde{F}^{\mu\nu}=\f{1}{2}\epsilon^{\mu\nu\rho\sigma}F_{\rho\sigma}$.
This action reduces to that of the Maxwell in the limit
$\beta\rightarrow\infty$. The field equation for the BI field is
obtained from the above Lagrangian
\begin{align}
 \nabla_\mu P^{\mu\nu}=0,\label{fef1}
\end{align}
where we have introduced the second rank tensor $P^{\mu\nu}$
\begin{align}
 P^{\mu\nu}=-\f{1}{2}\f{\partial L(F)}{\partial F_{\mu\nu}}=
\f{F^{\mu\nu}-\f{1}{4\beta^2}\left(\tilde{F}^{\rho\sigma}F_{\rho\sigma}\right)\tilde{F}^{\mu\nu}}
{\mc{R}}. \label{fef2}
\end{align}
We note that in the weak field limit $P^{\mu\nu}\approx F^{\mu\nu}$.
The energy-momentum tensor can be written as
\begin{align}
\tau_{\mu\nu}=\f{1}{4\pi}\left[\f{F_\mu^{~\lambda}F_{\nu\lambda}+\beta^2\left(
\mc{R}-1-\f{1}{2\beta^2}F^{\rho\sigma}F_{\rho\sigma}\right)g_{\mu\nu}}
{\mc{R}}\right].\label{feq1}
\end{align}
In general, the anti-symmetric tensor $F_{\mu\nu}$ has six
independent components. However, since the Einstein and $Q_{\mu\nu}$
tensors are diagonal, the six off-diagonal components of (\ref{eq1})
result in six algebraic equations with solutions where only two
non-zero components, $F^{tx}$ and $F^{yz}$, survive. Using the field
equation for the BI matter (\ref{fef1}) and  conservation equation
for the BI energy-momentum tensor, $\nabla^\mu\tau_{\mu\nu}=0$, one
can find the  two remaining components
\begin{align}
 F^{tx}=\f{\beta f_1}{\sqrt{16\beta^2 v^2+a_1^2}},\qquad F^{yz}=-\f{f_2}{4}\left(\f{a_1}{v}\right)^2, \label{comp}
\end{align}
where $f_i$ are constants and
\begin{align}
 f_1^2+f_2^2=1.\label{cons}
\end{align}
This relation ensures that the conservation equation and the field
equation of the BI matter are satisfied and eliminates the constants from
the energy-momentum tensor which can be written using equation
(\ref{feq1}) as
\begin{align}
 &\tau^0_{~0}=\tau^1_{~1}=\f{\beta^2}{4\pi}\left(1-\Delta\right), \nonumber \\
&\tau^2_{~2}=\tau^3_{~3}=\f{\beta^2}{4\pi}\left(1-\f{1}{\Delta}\right),
\label{en2}
\end{align}
where
\begin{align}
 \Delta=\sqrt{1+\f{1}{16\beta^2}\left(\f{a_1}{v}\right)^2}.\label{delta}
\end{align}
The energy-momentum tensor of the BI field can be written in the
form of a perfect fluid $\tau^\mu_\nu=
\textmd{diag}(-\rho,p^{\parallel},p^\perp,p^\perp)$ with the
equation of state of the form $p^{\parallel}=-\rho$. Another
equation of state relating the transverse pressure can be obtained
from (\ref{en2})
\begin{align}
 p^{\perp}=\f{\rho}{1-\f{4\pi}{\beta^2}\rho}. \label{eqs}
\end{align}
In the limit $\beta\rightarrow\infty$ one can see from equation
(\ref{eqs}) that the energy-momentum tensor of the BI field takes
the form $\textmd{diag}(-\rho,-\rho,\rho,\rho)$. As can be expected,
this is the energy-momentum of the Maxwell's field with
\begin{align}
 \rho_{_{Maxwell}}\propto\left(\f{a_1}{v}\right)^2,\label{max}
\end{align}
where $v$ is defined in (\ref{def1}). Equations (\ref{ext22}),
(\ref{en2}) and the Einstein tensor for the Bianchi I metric suggest
that the scale factors in the $(yy)$ and $(zz)$ directions are
proportional. Solving the $(yy)$ and $(zz)$ components of the field
equation (\ref{eq1}) one obtains
\begin{align}
\f{a_3(t)}{c_3}=\f{a_2(t)}{c_2}. \label{equal}
\end{align}
This means that the second and third components of the Einstein
equation are identical. Equation (\ref{equal}) also implies that the
constants $c_2$ and $c_3$ must have the same sign. It only remains
to solve the field equations (\ref{eq1}) which, using equation
(\ref{def1}) can be written as
\begin{widetext}
\begin{align}
H_2^2+2H_1
H_2-\Lambda-3+2\f{c_1}{a_1}+4\f{c_2}{a_2}-2\f{c_1}{a_1}\f{c_2}{a_2}
-\left(\f{c_2}{a_2}\right)^2=\f{\alpha\beta^2}{4\pi}\left(\Delta-1\right),\label{eqf1}
\end{align}
\begin{align}
3H_2^2+2\dot{H}_2-\Lambda-3+4\f{c_2}{a_2}-\left(\f{c_2}{a_2}\right)^2=\f{\alpha\beta^2}{4\pi}\left(\Delta-1\right)\label{eqf2},
\end{align}
\begin{align}
H_1^2+H_2^2+H_1 H_2+\dot{H_1}+\dot{H_2}-\Lambda-3+2\f{c_1}{a_1}
+2\f{c_2}{a_2}-\f{c_1}{a_1}\f{c_2}{a_2}=\f{\alpha\beta^2}{4\pi}\left(\f{1}{\Delta}-1\right),\label{eqf3}
\end{align}
where $\Delta$ takes the form
\begin{align}
 \Delta=\sqrt{1+\left(\f{c_2}{4\beta c_3 a_2^2}\right)^2}.\label{newdelta}
\end{align}
Equation (\ref{eqf2}) can be integrated to find the scale factor
$a_2$ implicitly
\begin{align}
 t=\int\f{\mathrm{d}a_2}{\sqrt{\left(1+\f{\Lambda}{3}\right)a_2^2-\left(\f{\alpha\beta^2}{12\pi}+2c_2\right)a_2+c_2^2
+\f{c_4}{a_2}-\f{\alpha\beta^2}{4\pi}\f{1}{a_2}\int^{a_2}\sqrt{f^4+\left(\f{c_2}{4\beta
c_3}\right)^2}\mathrm{d}f}}.\label{bb}
\end{align}
\end{widetext}
One can also find a relation between the scale factors $a_1(t)$ and
$a_2(t)$
\begin{align}
 a_1=\dot{a_2}\left(c_1\int\f{c_2-a_2}{\dot{a}_2^2}\textmd{d}t-c_5\right),
\label{aa}
\end{align}
where $c_5$ is an integration constant.
\section{Late time behavior}
An explicit form for the solution of the above equations seems to be
unavailable. It is therefore desirable to analyze the asymptotic
behavior of the observable quantities at late times. Assuming
$\beta<\infty$, the behavior of the energy-density and pressure of
the BI field is found to be
\begin{align}
\rho&=\f{c_2^2}{128\pi c_3^2}\f{1}{a_2^4}-\f{c_2^4}{8192\pi c_3^4\beta^2}\f{1}{a_2^8}+\mathcal{O}\left(\f{1}{a_2^{12}}\right),
\nonumber\\
p^\perp&=\f{c_2^2}{128\pi c_3^2}\f{1}{a_2^4}-\f{3c_2^4}{8192\pi
c_3^4\beta^2}\f{1}{a_2^8}+\mathcal{O}\left(\f{1}{a_2^{12}}\right).
\label{asy1}
\end{align}
Equations (\ref{asy1}) are obtained by writing $H_2$ in terms of
$a_2$ from equation (\ref{bb}) followed by writing $H_1$ in terms of
$a_2$ using equation (\ref{aa}). These equations
show that at late times the BI field differ from that of the Maxwell
in the second terms which is of order $a_2^{-8}$. Consequently, the
BI field very rapidly turns into the Maxwell field unless the
constant $\f{c_2^4}{c_3^4\beta^2}$ is very large. One can also see
that the anisotropy  and deceleration parameters behave according to
\begin{align}
 A&=\f{A_2}{a_2^2}+\mathcal{O}\left(\f{1}{a_2^3}\right),\label{Aa}\\
 q&=-1+\f{q_1}{a_2}+\f{q_2}{a_2^2}+\mathcal{O}\left(\f{1}{a_2^3}\right),\label{Qq}
\end{align}
where we have defined
\begin{align}
& A_2=\f{6}{c_5^2}\f{\left(c_1-\f{1}{3}c_2c_5\sqrt{9+3\Lambda}\right)^2}{\left(\Lambda+3\right)^3},\nonumber\\
& q_1=\f{2c_2c_5(\Lambda+3)-c_1\sqrt{3\Lambda+9}}{c_5(\Lambda+3)^3},\nonumber\\
&
q_2=\f{17\sqrt{9-3c_1c_2c_5\Lambda}+c_2^2c_5^2(\Lambda+3)(\Lambda-2)+12c_1^2}{c_5^2(\Lambda+3)^3}.
\end{align}
\begin{figure}
\includegraphics[scale=1]{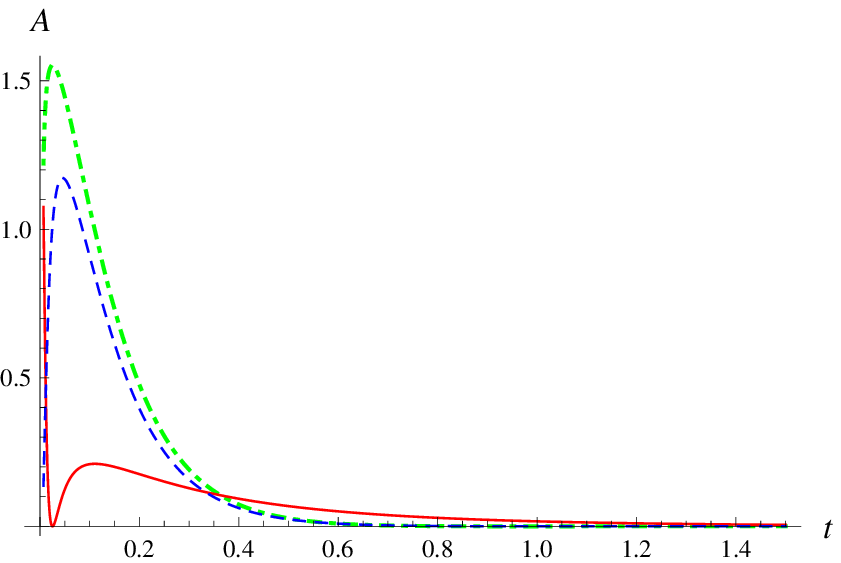}
\centering \caption{The Anisotropy parameter in the cases:
$\beta=\infty$ (dashed-dotted line), $\beta=1$ (dashed line),
$\beta=10^{-5}$ (solid line)} \label{fig1}
\includegraphics[scale=1]{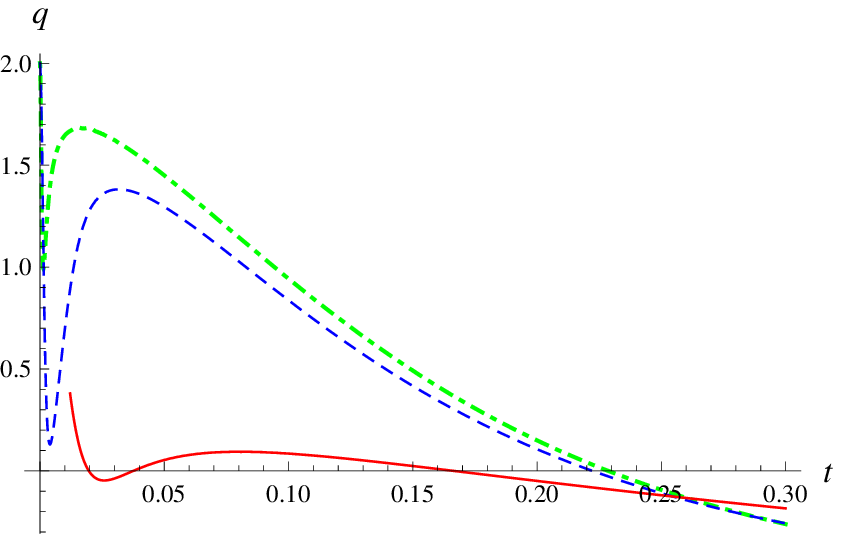}
\caption{The early time behavior of the deceleration parameter in
the cases: $\beta=\infty$ (dashed-dotted line), $\beta=1$ (dashed
line), $\beta=10^{-5}$ (solid)} \label{fig2}
\includegraphics[scale=1]{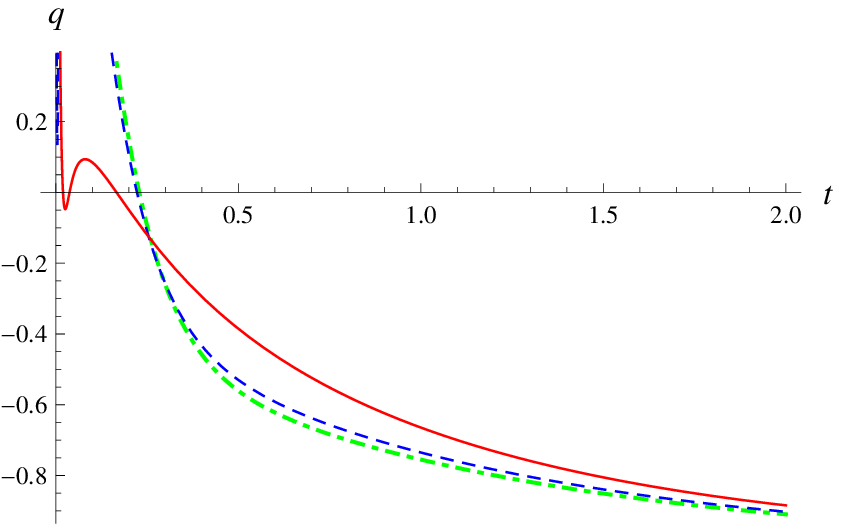}
\caption{The late time behavior of the deceleration parameter in the
cases: $\beta=\infty$ (dashed-dotted line), $\beta=1$ (dashed line),
$\beta=10^{-5}$ (solid)} \label{fig3}
\end{figure}
The asymptotic behavior of the scale factor is
\begin{align}
 a_1\sim \f{c_5}{3}\sqrt{3\Lambda+9}a_2. \label{appro}
\end{align}
One can see from  equations (\ref{equal}) and (\ref{appro}) that the
Hubble parameter of the three scale factors become equal at late
times, predicting an isotropic universe. In figures (\ref{fig1}),
(\ref{fig2}) and (\ref{fig3}) we have plotted the anisotropy and
deceleration parameters for three different values of $\beta$,
including the Maxwell's regime $\beta=\infty$ and  constants
$c_1=-100,\,c_2=-1,\,c_3=-60$. These figures show that the behavior
of the anisotropy parameter and the deceleration parameter are
largely dependent on the Born-Infeld coupling. As the constant
$\beta$ decreases, the early time anisotropy of the BI universe
becomes less pronounced. From the first figure we see that the slope
of the anisotropy parameter decreases, leading to a less anisotropic
universe which becomes isotropic with a smaller rate than the
Maxwell universe. For $\beta=10^{-5}$, the universe reaches an
isotropic stage at early times, becoming less anisotropic again and
finally reaches the present isotropic state. The deceleration
parameter diagram shows the same behavior; the BI universe starts
from a smaller amount of deceleration at early times and becomes
accelerating considerably slower than the Maxwell case at late
times.

To answer the question of what value of $\beta$ is the most
suitable, one requires comparison against the observational data
which may become a realistic possibility in not too distant a
future. In figure (\ref{fig2}) we have plotted the early time
behavior of the deceleration parameter. One can see from the figure
that the BI universe with the coupling $\beta=10^{-5}$ has an early
accelerating stage which alternately becomes decelerating and
accelerating at late times. Finally we note that the late time
behavior of the theory discussed in this section does not depend on
the values of the constants of the model. Indeed at late times the
model asymptotes the Maxwell's theory, which can be seen from the
figures. However the detailed early time behavior of the model
depends on the values of the constants of the model, but its
qualitative behavior is the same as the constants assume different
values.
\section{Conclusions and Remarks}
The observation of the present acceleration of the universe is one
of the most intriguing  problems in the present-day cosmology.
Explanations based on the existence of dark energy or modified
theories of gravity are common approaches to describe it.

In this paper we have dealt with this problem in the context of a
brane-world model in which the assumption of the $\mathbb{Z}_2$
symmetry and use of the Israel junction conditions are relaxed and
the matter content of the universe is assumed to be of the BI type.
To calculate the conserved quantity $Q_{\mu\nu}$ appearing in the
field equations, one needs to calculate the components of the
extrinsic curvature. We have used the Codazzi equation to achieve
this. The components of the extrinsic curvature are obtained by the
choice of an ansatz for its (00) component. However, as long as the
bulk metric is not specified, this component can be chosen in such a
way as to make the tensor $Q_{\mu\nu}$ taking as simple a form as
possible. In the SMS method, this is achieved by the form of the
energy-momentum tensor. However, as is discussed in the Appendix,
the form of the BI energy-momentum tensor is not arbitrary and is
calculated from the BI field equations, indicating that the SMS
procedure may not provide solutions compatible with the present
observations when confining the BI field as the matter source to the
brane. In contrast, the method used in this paper provides more
freedom in choosing the extrinsic curvature, making the confinement
of the BI matter on the brane a possibility. In other words, in the
SMS method one restricts the bulk geometry to be satisfied by the
Israel junction conditions, leaving the energy-momentum tensor
arbitrary. In our model however, the procedure does not have such a
restriction.

The asymptotic behavior of the BI matter was also studied. As can be
seen from equation (\ref{asy1}), its behavior mimics that of the
Maxwell field in a certain limit; the BI field makes the transition
of the universe  from a highly anisotropic early state to its
present isotropic form slower than the Maxwell field and this
depends on the parameter $\beta$, as is seen from figure
(\ref{fig1}).

\section{Appendix}
Let us now discuss the consequences of using the BI field in  brane
world scenarios where the confinement of gauge fields is achieved
through the SMS method. As is well known, in this method one uses
the Israel junction conditions to replace the extrinsic curvature
tensor with the energy-momentum tensor. It follows that a new
tensor, $\pi^{\mu\nu}$, emerges which is quadratic in the
energy-momentum tensor. This tensor can be written as
\begin{align}
 \pi^{\mu\nu}=-\f{1}{4}\tau_{\mu\alpha}\tau_\nu^{~\alpha}+\f{1}{12}\tau\tau_{\mu\nu}+
 \f{1}{8}g_{\mu\nu}\tau_{\alpha\beta}
\tau^{\alpha\beta}-\f{1}{24}g_{\mu\nu}\tau^2,\label{pi}
\end{align}
where $\tau=\tau^\alpha_{~\alpha}$ and $g_{\mu\nu}$ is the brane
metric. One of the consequences of the SMS field equations is that
for the constant curvature bulk which we have considered in this
paper, the above tensor must be conserved $\nabla_\mu\pi^{\mu\nu}=0$
\cite{12}. If we consider the Bianchi I metric and note that the
energy-momentum tensor for this space-time takes the form
$\tau^\mu_\nu=\textmd{diag}(-\rho,-\rho,p,p)$, equation (\ref{en2}),
the conservation equation for $\pi^{\mu\nu}$ reduces to
\begin{align}
(\rho+p)(\dot{\rho}+2\dot{p})=0. \label{eqr}
\end{align}
The first term cannot be zero as one can see easily from definitions
of $\rho=\tau_0^0$ and $p=\tau_2^2$ from equations (\ref{en2}). So
the only equation of state for the BI matter, consistent with the
above constraint is $\rho+2p=\mbox{const}$. This equation can be
solved for the scale factors, with  solution
\begin{align}
 a_2\propto\f{1}{a_3}.\label{catas}
\end{align}
This solution is obviously ruled out by present observations.


\end{document}